\documentclass[aps,prd,superscriptaddress,nofootinbib,preprint]{revtex4}

\usepackage{float}
\usepackage{amsmath,amssymb}
\usepackage[dvips]{graphicx}
\DeclareGraphicsExtensions{.jpg,.mps,.pdf,.png,.eps} 
\usepackage{color}

\newcommand{\beq}{\begin{equation}}
\newcommand{\eeq}{\end{equation}}

\newcommand{\be}{\begin{equation}}
\newcommand{\ee}{\end{equation}}

\newcommand{\bea}{\begin{eqnarray}}
\newcommand{\eea}{\end{eqnarray}}
\newcommand{\non}{\nonumber}

\begin{document}

\title{Inflation and reheating in theories with spontaneous scale invariance symmetry breaking.}

\author{Massimiliano Rinaldi}
\email{massimiliano.rinaldi@unitn.it}
\author{Luciano Vanzo}
\email{luciano.vanzo@unitn.it}
\affiliation{Department of Physics, University of Trento\\ Via Sommarive 14, 38123 Trento, Italy}
\affiliation{INFN - TIFPA \\ Via Sommarive 14, 38123 Trento, Italy }


\begin{abstract} 
\noindent We study a scale-invariant model of quadratic gravity with a non-minimally coupled scalar field. We focus on cosmological solutions and find that scale invariance is spontaneously broken and  a mass scale naturally emerges. Before the symmetry breaking, the Universe undergoes an inflationary expansion with  nearly the same observational predictions  of  Starobinsky's model. At the end of inflation, the Hubble parameter and the scalar field converge to a stable fixed point through damped oscillations and the usual Einstein-Hilbert action is recovered. The oscillations around the fixed point can reheat the Universe in various ways and we study in detail some of these possibilities.
\end{abstract}

 \maketitle

 
 \section{Introduction}

 
\noindent The wealth of recent  observational data has dramatically reduced the number of viable inflationary models and opened a debate on the possibilities that are left \cite{oslo}. For instance,  in the context of single-field inflation, the data from Planck  \cite{planck15} basically exclude all potentials of the type $V\sim \phi^{p}$ with $p\geq 2$, thus leaving only a handful of feasible alternatives. Among these, there is the  Starobinsky model \cite{staro}, which is based on a minimal extension of general relativity obtained by the addition of a term quadratic in the Ricci scalar $R$. Originally motivated by quantum corrections, this model has been generalised to the class of so-called $f(R)$ gravity (see \cite{defelice,faraoni,capoz} for comprehensive reviews).

The  term  $R^{2}$ of the Starobinsky model  is dominant over the linear term during inflation and this reflects the idea that, at very high energy, gravity is fundamentally scale-invariant. In fact, the vacuum equations of motion obtained from the scale-invariant Lagrangian $L\sim \sqrt{g}R^{2}$ with a Robertson-Walker metric have a general solution that interpolates between an unstable radiation-dominated Universe (with $R=0$) and a stable de Sitter solution (with $R=$ const). The addition of a term proportional to $R$ breaks the scale invariance and introduces the Planck mass together with another mass scale of the order $10^{13}$ GeV. It also changes the stability properties of the solution in such a way that the Universe now evolves from an unstable de Sitter solution towards an oscillating phase that opens the door to reheating via parametric amplification of the field content of the theory. The Starobinsky model has become increasingly appealing  because of its conceptual simplicity, its close connection to Higgs inflation \cite{higgs} and, above all, because the predicted spectral indices are fully compatible with observations. In particular, the model predicts a tensor-to-scalar ratio of the order of $r\simeq 0.003$, which is well inside the upper limit set by Planck ($r<0.1$) and other experiments, such as the combined Keck and BICEP2 data ($r<0.07$) \cite{bicep}.

In this paper we propose a model where the scale invariance is broken dynamically. There is only one mass scale that  emerges as the value, at a stable fixed point, of a fundamental scalar field non-minimally coupled to gravity, but the precise value of the emerging mass scale is not determined. One of the advantages of this model is that there is no need of a second mass scale, as in the Starobinsky model, but the predictions for the spectral indices are essentially the same. In addition, in the broken phase, a residual cosmological constant arises. The latter can be related to the current vacuum energy in the Universe, which leads however to unnaturally small parameters (for example, the self-coupling of the quartic scalar interaction would take the unnatural value of order $10^{-122}$). Therefore the Hubble constant at the broken phase must be  considered as the initial data for the following radiation-dominated era. From a conceptual point of few, it is also important to stress that the model assumes scale-invariance as a fundamental symmetry of the system, which helps to restrict the form of Lagrangian density out of a very large number of possibilities.

The present paper is also motivated by the renewed interest in scale-invariant models of gravity in the recent years. To begin with, it has been shown that the most likely form of $f(R)$, in the absence of matter, compatible with the measured spectral indices is $f\sim R^{2-\delta}$, with $0<\delta\ll 1$ \cite{tn1}, see also \cite{reconstructing}. Other studies showed that scale-invariant gravity with quantum corrections  can reproduce inflation and spectral indices in line with current observations \cite{tn2} in the absence of inflaton or any other kind of matter fluid. Quantum corrections  in the context of inflation have been recently considered in various approaches, see e.g. \cite{agravity, effective, asafe,sannino}. Recently, a lot of work is being devoted to the so-called $\alpha$-attractors. This model, motivated by supergravity, provides a unified description of several inflationary models  by means of a unique parameter $\alpha$ related to the analicity properties of the scalar potential \cite{attractors}. It was shown that there exists another  class of attractors that are somewhat orthogonal to these and that have the fundamental property of being quasi scale-invariant \cite{tn3}, in contrast to the $\alpha$-attractors. Finally, scale-invariant gravity has been investigated also in the context of black hole physics where interesting thermodynamical properties were found \cite{R2bh}. We emphasise throughout the paper the role of scale symmetry (or dilatations) as a global one, in contrast to local conformal symmetry in the presence of a dynamical metric which, being dependent on an arbitrary function is really to be considered as a gauge symmetry, as amply explained in \cite{Hertzberg:2014aha}.

The plan of the paper is the following. In the next section we present the model, its symmetries and the equations of motion. In sec.\ \ref{sec3} we study the global dynamics, the fixed points and their stability, both analytically and numerically. In sec.\ \ref{sec4} we study the inflationary phase in the Jordan frame. In sec.\ \ref{sec5} we present various reheating mechanisms that can be applied to this model. In sec.\ \ref{sec6} we review the inflationary phase in the Einstein frame and we prove that the predicted spectral indices are the same as in the Starobinsky model, at least at the leading order. We finally conclude in sec.\ \ref{sec7} with some considerations.


 \section{The Lagrangian and its symmetries}\label{sec2}

\noindent Our  model is based on the scale-invariant Lagrangian 
\bea\label{fulllagra}
{\cal L}_{\rm inv}=\sqrt{|\det g|}\left[ {\alpha\over 36}R^{2}+{\xi\over 6}\phi^{2} R-{1\over 2}(\partial\phi)^{2}-{\lambda\over 4}\phi^{4}\right]\,,
\eea
where $\alpha$, $\lambda$, and $\xi$ are  positive constant. Normally, one should also add the standard model Lagrangian ${\cal L}_{SM}$ although, in most theories of inflation, and in our model as well, it can be omitted except, eventually, for the  Higgs field.  Note that scale invariance forbids the appearance of a cosmological constant term, although such a term may well appear after the breaking of the scale symmetry. The field $\phi$ is prevented to interact with the standard model fields due to the $SU(3)\times SU(2)\times U(1)$ gauge symmetry (except for the Higgs and right-handed neutrinos) but it couples to them indirectly via the metric, a fact possibly relevant for the reheating phase. Also, for the conformally flat FRWL background considered here, the term $R^2$ is the only one surviving the conformally flat space limit since the Weyl term $W_{\alpha\beta\mu\nu}W^{\alpha\beta\mu\nu}$ vanishes there. We note that if  $\alpha$ were to vanish and $\xi=1$ the theory would be conformal invariant, but not equivalent to standard general relativity with a $\Lambda$-term, as would be the case if the scalar kinetic terms had the opposite sign. We also remark that scale invariance alone (as well as local Weyl invariance) cannot restrict the form of the counter-terms in a perturbative expansion around flat space, as has been suggested in the past \cite{Kallosh}, since quantisation necessarily breaks it. The fact remains that the standard model Lagrangian without the Higgs field but with all the fermions massless is obviously scale invariant and could  also be added to \eqref{fulllagra}, although we shall not do this for the time being.  

Scale-invariance is manifest since the action is invariant\footnote{The Lagrangian is not invariant since it transforms as a density.} under the following active transformation laws\footnote{A field with canonical dimension $d$ is usually assumed to transform as $\phi'(x)=L^d\phi(L x)$ under $x'=L^{-1}x$.}
\bea\label{scinv}
\bar g_{\mu\nu}(x) = g_{\mu\nu}(\ell x)\,,\quad \bar \phi(x) = \ell \phi(\ell x)\,,
\eea
for any constant $\ell$, which we call dilatation symmetry. There is also a rigid internal Weyl symmetry, with parameter $L$, under which coordinates do not change,  $g'_{\mu\nu}(x)=L^2g_{\mu\nu}(x)$, $\phi'(x)=L^{-1}\phi(x)$, and which leaves the Lagrangian density strictly invariant. Finally there is the  product symmetry under which $g'_{\mu\nu}(x)=L^2g_{\mu\nu}(\ell x)$, $\phi'(x)=L^{-1}\ell\phi(\ell x)$. The special case $L=\ell$ is the usual diff symmetry $x^{\mu}\to\ell^{-1} x^{\mu}$ under which $\phi(x)$ transform as a scalar field and $g_{\mu\nu}(x)$ as a tensor field, and it is never broken. As a result, if the action is invariant, say, under the Weyl transformation then it is also invariant under dilatations, and viceversa. We emphasise, however, that there really is a two-parameter ($L$ and $\ell$) abelian group of symmetries, although one cannot break one of these without breaking the other unless we also break the diff symmetry\footnote{Precisely because we have a two-parameter group we can break scale symmetry without spoiling diffeomorphism invariance.}. 

The derivative with respect to $\phi$, with $R$ fixed, of the effective potential
\bea
V_{\rm eff}=-{\xi\over 6}\phi^{2} R+{\lambda\over 4}\phi^{4}\,,
\eea
vanishes at
\bea
\phi=0\,,\qquad \phi^{2}_{0}={\xi R\over 3\lambda}\,,
\eea
the first point being a local maximum and the second a local minimum. Thus, in principle there can be a classical symmetry breaking of the scale symmetry (and the discrete symmetry $\phi\to-\phi$ as well,  possibly leading to a domain wall structure) that occurs when the scalar field settles in the minimum over some infinite volume region of space-time with constant curvature. This automatically introduces in the theory a mass scale that can be identified with $\phi_{0}$. A similar mechanism was studied in \cite{bellido}, where the quadratic term in $R$ is replaced by another dynamical scalar field, and in a more general context including conformal invariance by Bars et al. \cite{Bars:2013yba}. All these models are also inspired by induced gravity models, see e.g. \cite{Venturi}.

For $\alpha=\xi^2/\lambda$, the non derivative part of the Lagrangian density \eqref{fulllagra} takes the  form
\bea\label{effep}
-\sqrt{|\det g|}\,\frac{\lambda}{4}\left(\phi^2-\frac{\xi R}{3\lambda}\right)^2\,.
\eea
which vanishes at the minimum (see also Eq.~\eqref{alpha} below). Note that all minima connected by a scale transformation (under the Weyl or product symmetry) have the same vanishing energy. Therefore, the potential has flat directions corresponding to constant  Weyl deformations of the the scalar field and of the metric. By defining $\phi=M\exp(\sigma/M)$ and $g_{\mu\nu}(x)=\exp(-2\sigma/M)\tilde{g}_{\mu\nu}(x)$, for some mass scale $M$, we see that the interactions could only depends on $\sigma$ via derivative terms since, for constant $\sigma$, the Lagrangian density is invariant. The field $\sigma$ is the dilaton of the theory, namely the Goldstone boson associated with the breaking of the rigid Weyl symmetry, and it transforms non-linearly as $\sigma\to\sigma+M\log\ell$. In fact,  it generates a Einstein-Hilbert coupling $M^2\xi R/6$, that defines  the effective Newton constant (see, for example \cite{Demir:2004kc}). It should be said that in the logic of effective field theory there is clearly room for infinitely many more scale invariant terms in the Lagrangian containing increasing powers of derivatives, but we stick here to the lowest order dominant terms. Having discussed the role  of the field $\phi$ as a dilaton, we  now study the equations of motion on a cosmological background.

We choose a flat Robertson-Walker (RW) metric of the form $ds^{2}=-dt^{2}+a^{2}(t)\delta_{ij}dx^{i}dx^{j}$. The Hubble parameter is defined as $H=d\ln a/dt$ and the Ricci scalar is $R=12H^{2}+6\dot H$. There are two independent equations of motion that read
\bea\label{eomt}
&&\ddot\phi+3H\dot \phi-2\xi\phi\dot H-\phi(4\xi H^{2}-\lambda\phi^{2})=0\,,\\\non
&& \alpha\left(2H\ddot H  -\dot H^{2}+6H^{2}\dot H  \right)-{1\over 2}\dot\phi^{2}+2\xi\phi \dot \phi H+{\phi^{2}\over 4}(4\xi H^{2}-\lambda \phi^{2})=0\,.
\eea
Scale invariance now takes the form of invariance under the rescaling to new fields defined by
\bea\label{scinv2}
\bar \phi(t) = \ell \phi (\ell t)\,,\quad \bar a(t)=a(\ell t)\,,\quad \bar H(t) =\frac{1}{\bar a}\frac{d{\bar a}}{dt}=\ell H(\ell t)\,,
\eea
for an arbitrary $\ell$ that leaves these equations unchanged\footnote{They are not invariant under rigid Weyl transformations because, by choosing the Robertson-Walker metric, we have gauge-fixed the diff symmetry.}.
For future calculations, it is  convenient to write eqs.\ \eqref{eomt} in terms of the e-folding time $N=\ln a$ 
\bea\label{eomN}
&&H^{2}\phi''+(HH'+3H^{2})\phi'-2\xi\phi HH'-\phi(4\xi H^{2}-\lambda\phi^{2})=0\,,\non\\
&&\alpha H^{2}\left(2HH''+H'^{2}+6HH'\right)+2\xi H^{2} \phi \phi'-{1\over 2}\phi'^{2}H^{2}+{\phi^{2}\over 4}(4\xi H^{2}-\lambda \phi^{2})=0\,.
\eea
Here, the prime stands for a the derivative with respect to $N$. 

The model has three free parameters $(\alpha,\xi,\lambda)$ but in fact one can be eliminated by requiring that, when both $H$ and $\phi$ are constant, the quadratic term in $R$ and the quartic term in $\phi$ cancel each other, so that, in the  Lagrangian there is a vanishing cosmological constant in this regime. Thus, by setting $H=H_{0}$ and $\phi=\phi_{0}$ in eqs.\ \eqref{eomt}, we find that
\bea\label{zeros}
\phi_{0}=0\,,\quad {\rm or}\quad \phi_{0}=2H_{0}\sqrt{\xi\over \lambda}\,.
\eea
Then, by imposing the relation
\bea
{\lambda\phi_{0}^{4}\over 4}={\alpha R(H_{0})^{2}\over 36}\equiv 4\alpha H_{0}^{4}\,,
\eea
together with the second of the solutions \eqref{zeros}, we find that
\bea\label{alpha}
\alpha={\xi^{2}\over \lambda}\,,
\eea
a relation that will be adopted from now on. 
Note that the solutions \eqref{zeros} also coincide with the two extrema of this potential since, when $H$ is constant,  $R=12H^{2}$. Their existence displays the broken symmetry phase since the spatial volume of the flat RW metric is infinite. We anticipate that the two extrema corresponds also to the only two fixed points of the system of equations of motion. Therefore, if at least one of these is attractive and stable, the system will dynamically relax to one of the minimum of the effective potential \eqref{effep} realising a spontaneous breaking of the scale symmetry. But since any minimum is as good as any other, the values of the parameters at the fixed point will be undetermined. It should  be stressed that, were it not for the presence of the standard model fields, perturbations around the fixed point corresponding to the minimum of the potential would still obey scale invariant field equations, except that the scale symmetry would perhaps be realised non-linearly. In fact the symmetry is broken in the vacuum, not in the field equations (or in the Lagrangian).


\section{Global evolution}\label{sec3}


\noindent The most convenient way to analyse the global evolution of the two equations \eqref{eomN} is to convert them into a four-dimensional dynamical system, find its fixed points and study their stability.  We first use analytical methods to solve the system, linearised nearby the fixed points. We then verify these results by solving numerically the full equations.

\subsection{Fixed-point analysis}

\noindent To find the fixed points, we make the substitutions  
\bea\label{sys}
H(N)=x\,,\quad H'(N)=y\,,\quad \phi(N)=z\,,\quad \phi'(N)=w\,,
\eea
in eqs.\ \eqref{eomN} so that we write an equivalent system of four first-order differential equations that reads
\bea
&&w'+{yw\over x}+3w-{2\xi zy\over x}+{\lambda z^{3}\over x^{2}}-4\xi z=0\,,\\\non
&&y'+{y^{2}\over 2x}+3y+{\lambda wz\over \xi x}-{\lambda^{2} z^{4}\over 8 \xi^{2} x^{3}}+{\lambda z^{2}\over 2\xi x}-{\lambda w^{2}\over 4 x\xi^{2}}=0\,,\\\non
&&x'-y=0\,,\\\non
&&z'-w=0\,.
\eea
By solving the system $(w'=0,y'=0,x'=0,z'=0)$, we find two families of fixed points given by 
\bea
(x,y,z,w)=(x,0,0,0)\,,\quad (x,y,z,w)=\left(x,0,\pm 2\sqrt{\xi\over \lambda}\,x,0\right)\,,
\eea
for arbitrary $x$. Note that also the fixed points are scale-invariant under a redefinition of $x$ (i.e. of $H$).
By computing the Jacobian and the corresponding eigenvalues, we find that at least one is vanishing for both points, so we need to resort to analytical and numerical method to assess unambiguously the stability.

We begin by linearising the system around the point $(x,0,0,0,)$. The solution is
\bea\label{unsteq}
x&=&H(N)=c_{1}+c_{2}\,e^{-3N}\,,\\\non
z&=& \phi (N)=c_{3}\,e^{\left(-\frac32+\frac12 \sqrt{9+16\xi} \right)N}+c_{4}\,e^{\left(-\frac32-\frac12 \sqrt{9+16\xi} \right)N}\,,
\eea
where $c_{1\ldots4}$ are constants of integration. Since also $z'=w$ vanishes at the fixed point, if we impose that the latter belongs to the trajectory in the $(z,w)$ plane, then we are forced to set $c_{3}=c_{4}$. On the opposite, if both $c_{3}$ and $c_{4}$ are non-vanishing then we have a saddle point, as $z(N)$ is a combination of growing and decaying modes, since $\xi>0$ by hypothesis. We conclude that the point $(x,0,0,0)$ is a saddle point: the Hubble parameter tends to a constant while the scalar field grows dragging the system away from it.

Let us analyse the second fixed point, focussing on the positive solution $(x,0,2x\sqrt{\xi/ \lambda},0)$. In this case, the general solution to the linearised system is
\bea\label{oscil}
x&=&c_{1}+c_{2}e^{-3N}+e^{-{3\over 2}N}(c_{3}S(N)+c_{4}C(N))\,,\\\non
z&=&\sqrt{\xi\over \lambda}\left[2c_{1}+{c_{2}\over 2}\,e^{-3N}+{\xi\over 2(1+2\xi)}\,e^{-\frac32 N}\Big( (2Kc_{4}-5c_{3})S(N)-(5c_{4}+2Kc_{3})C(N) \Big)\right]\,,
\eea
where $K=\frac12 \sqrt{7+64\xi}$ and $S(N)=\sin (KN)$, $C(N)=\cos (KN)$. We clearly see that the fixed point is stable, it is an attractor that is reached through damped oscillations of both $H$ and $\phi$. 

The breaking of the scale invariance occurs when the the solution begins to oscillate around the stable fixed point and the sum of the quadratic term in $R$ and of the quartic term in $\phi$ vanishes through damped oscillations. At this stage,  the prefactor of the linear term in $R$ becomes constant and dimensionally equivalent to a mass scale. Thus, it is natural to make the identification
\bea
\frac16 \xi \phi_{0}^{2}R\equiv \frac12 M_{p}^{2}R\,,
\eea
from which, as anticipated above, we find that
\bea\label{mplanck}
M_{p}\equiv \sqrt{\xi\over 3}\,\phi_{0}\,.
\eea

The non-zero value of the Hubble parameter at the stable fixed point, say $H_{\star}$,  could in principle account for a fundamental non-vanishing cosmological constant. However,  this is true, in fact, only if the model that we have chosen is all there is, so $H_{\star}$ would persist for all time. According to the present understanding,  after inflation the Universe enters a radiation-dominated era during which the standard model Lagrangian that we omitted initially becomes important and scale invariance is also broken at the level of the Lagrangian (for example by the Higgs mass term). Therefore, it seems more sensible to think of $H_{\star}$ as the initial value of the Hubble parameter at the onset of the radiation-dominated era, whose value is valid at some precise epoch only. From there on it would start decreasing according to the standard cosmology. 

We can appreciate the difference between these two interpretations in a simple way. If $H_{\star}$ is related to a fundamental cosmological constant then in the infinite future, when all the matter content of the Universe is diluted away and oscillations are damped out, we can write the equality
 \bea
H_{\star}^{2}={\Lambda\over 3}\,,
\eea
where $\Lambda$ is the ``relic'' cosmological constant, which is of order $(10^{-42}\,\,\rm{Gev})^2$ . With the help of the equations above, we also find that
\bea
\Lambda={\lambda \phi_{0}^{4}\over 4M_{p}^{2}}\,.
\eea
So this requires a tremendous amount of fine tuning in $\lambda$ for any $\xi\sim 1$ \footnote{The choice $\xi\sim 1$ can be argued on the ground that $\xi=1/6$ gives a conformally invariant theory if the scalar field is a ghost, a situation that one can imagine to occur at even earlier times. Also, with $\xi\sim1$ one has $\phi_{0}\sim M_{p}$.}, as it is also apparent from the equivalent formula 
\bea\label{xilambda}
\xi={3M_{p}\over 2}\sqrt{\lambda\over \Lambda}\,.
\eea
Alternatively, we can express the parameter $\xi$ in terms of $\Lambda$ as
\bea
H_{\star}={M_{p}\sqrt{3\lambda}\over 2\xi}\,.
\eea
and treat $H_{\star}$ as an initial data. The scale of the Hubble parameter at inflation is roughly $H_{\star}\sim 10^{14}\,\,\rm{Gev}$, which gives,  assuming again $\xi\sim 1$, a coupling $\lambda\sim 10^{-8}$ (a weakly coupled scalar is good for inflation).  Furthermore, using the relation
\bea
H_{\star}=H_{0}\sqrt{\Omega_{m}(1+z_{\star})^3+\Omega_{r}(1+z_{\star})^4+\Omega_{v}}\simeq H_{0}\Omega_{r}^{1/2}(1+z_{\star})^2
\eea
where $H_{0}\simeq 10^{-42}\,\,\rm{Gev}$ is the present value of Hubble constant, we get the redshift $z_{\star}=10^{29}$ and further $67$ e-foldings from there to the present era.

To summarise, the picture that emerges from these findings is that in the solution space there are trajectories that connect an inflationary Universe to a graceful exit characterised by damped oscillations that can produce particles through standard model reheating mechanisms. In addition, the model has a ``residual'' Hubble parameter $H_{\star}$ which can either be interpreted as a residual cosmological constant, which would be a wrong interpretation, or as the initial data for the beginning of the radiation-dominated era. In that case the model does not explain the late time acceleration, but in the former case it requires a tremendous amount of fine tuning and an unnaturally small coupling parameter $\lambda$.

\subsection{Numerical analysis}

\begin{figure}[ht]
  \centering 
  \includegraphics[scale=0.39]{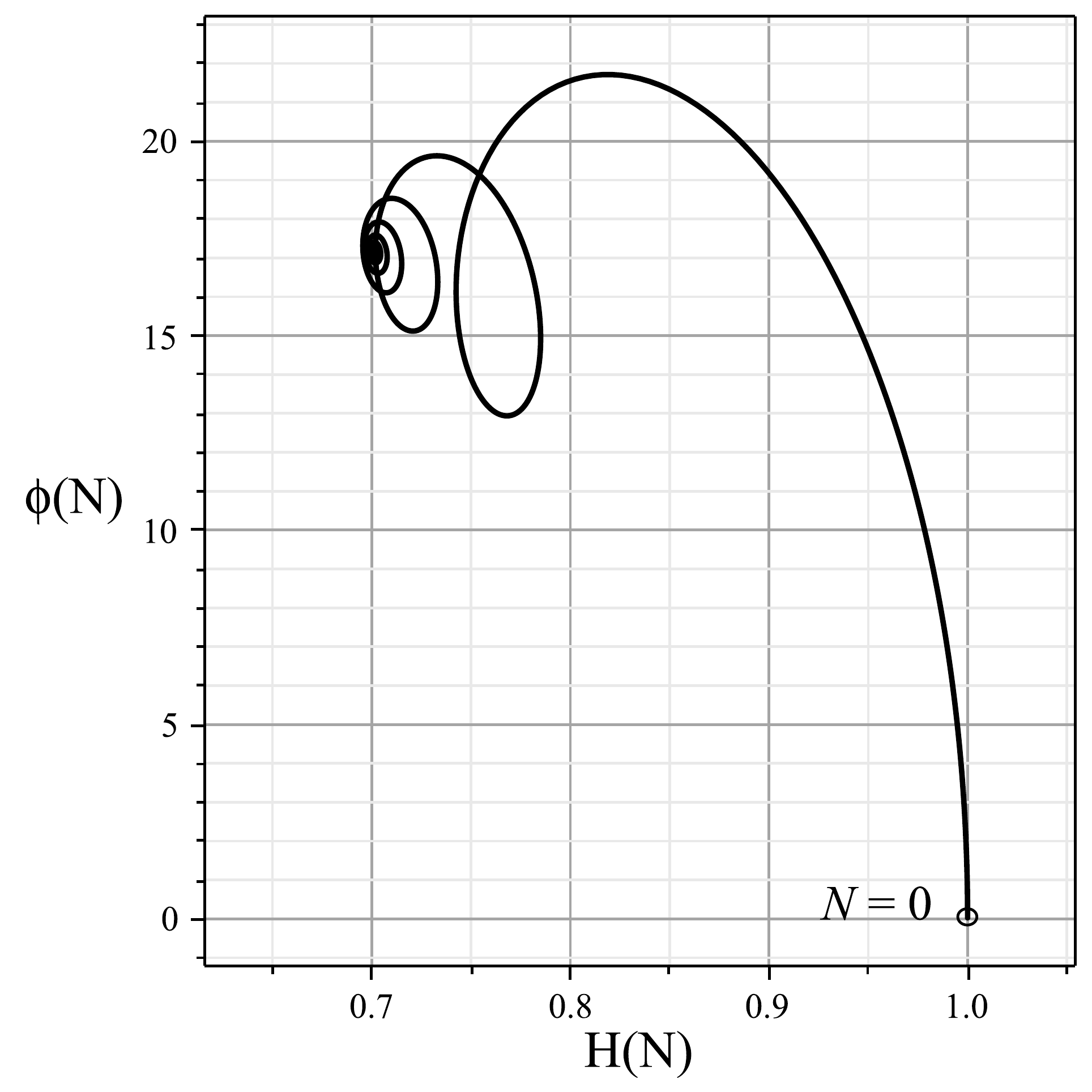} 
   \includegraphics[scale=0.39]{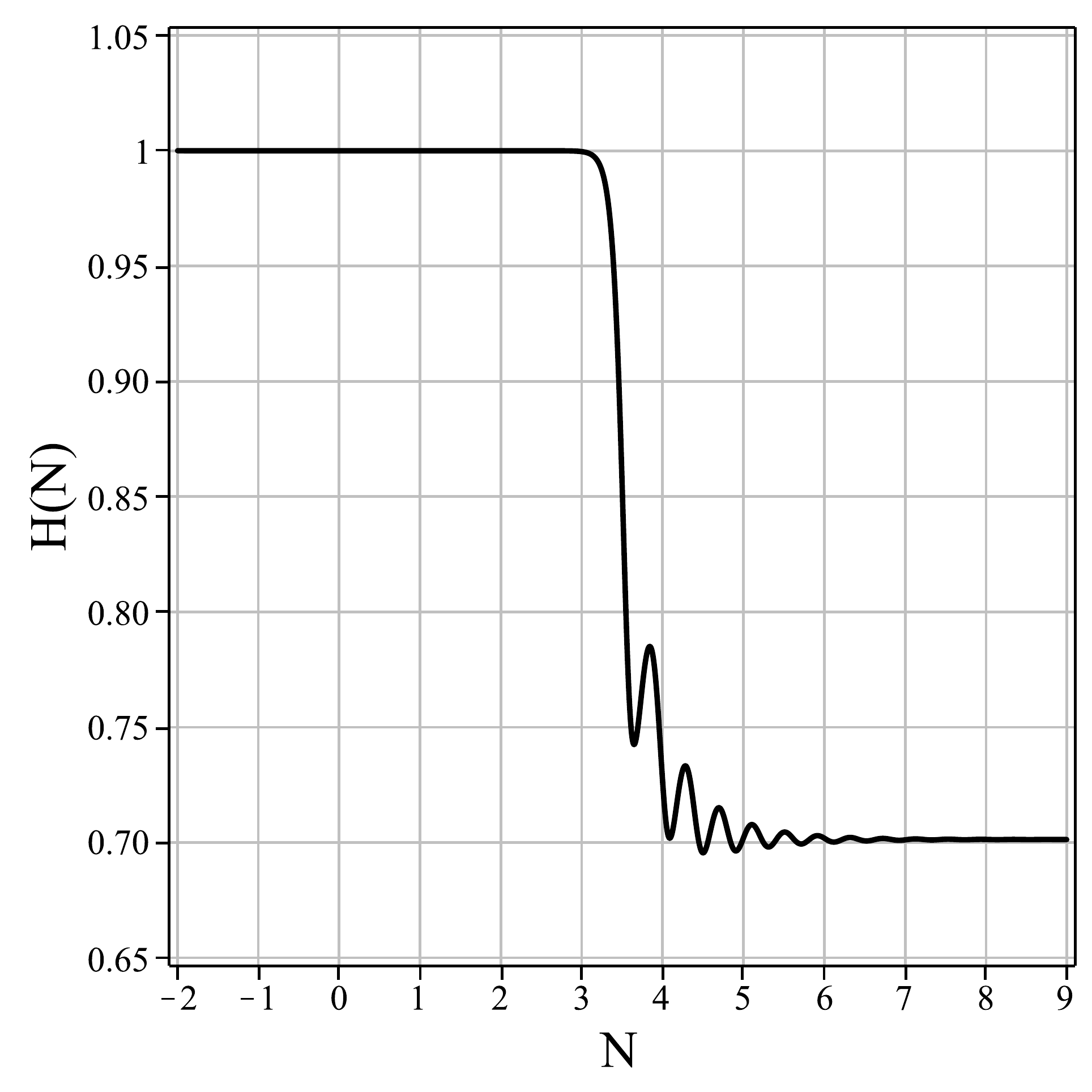} 
  \caption{Phase portrait of $\phi(N)$ and $H(N)$ (left) and plot of $H(N)$ (right).}
  \label{fig1}
  \end{figure}

\begin{figure}[ht]
  \centering 
  \includegraphics[scale=0.39]{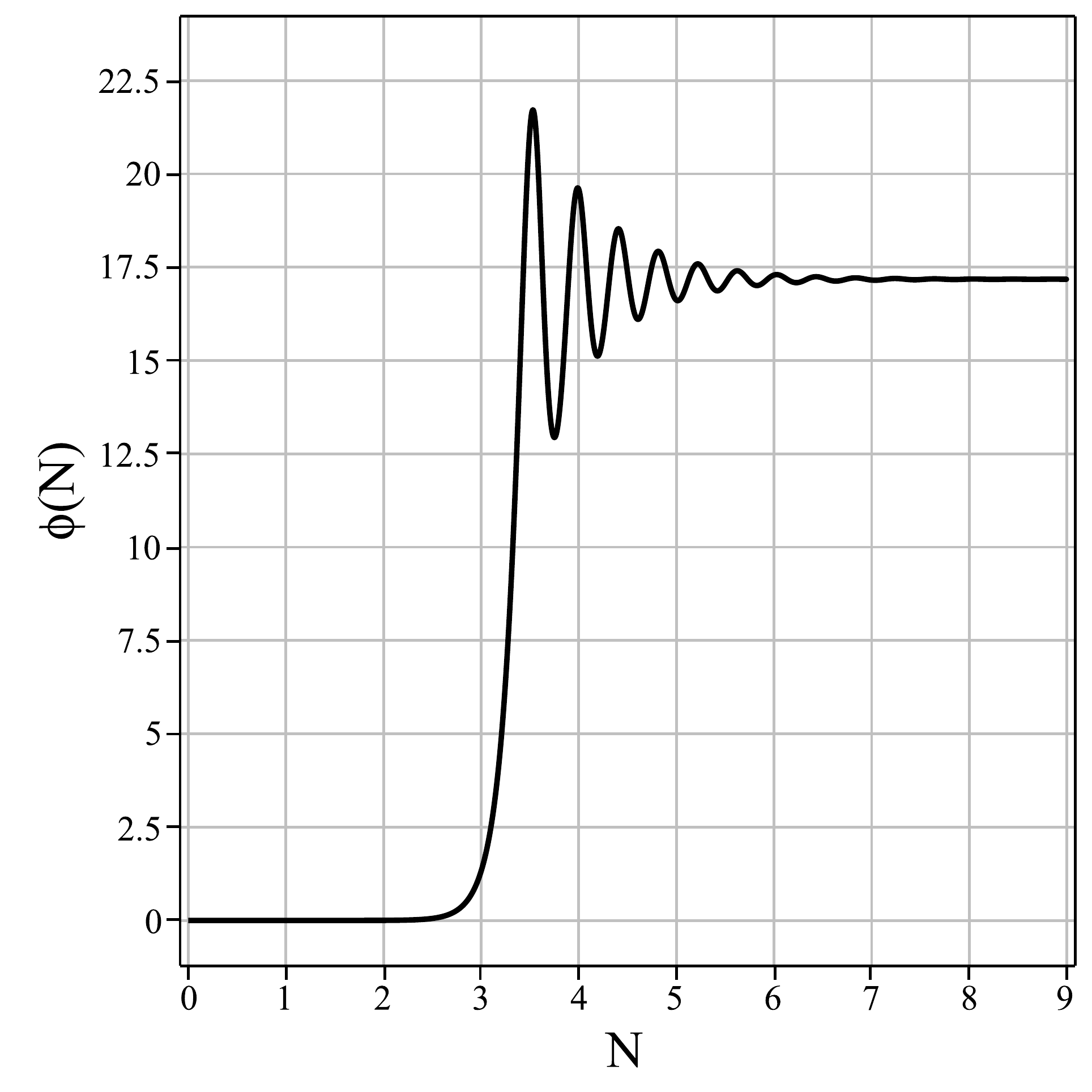} 
   \includegraphics[scale=0.39]{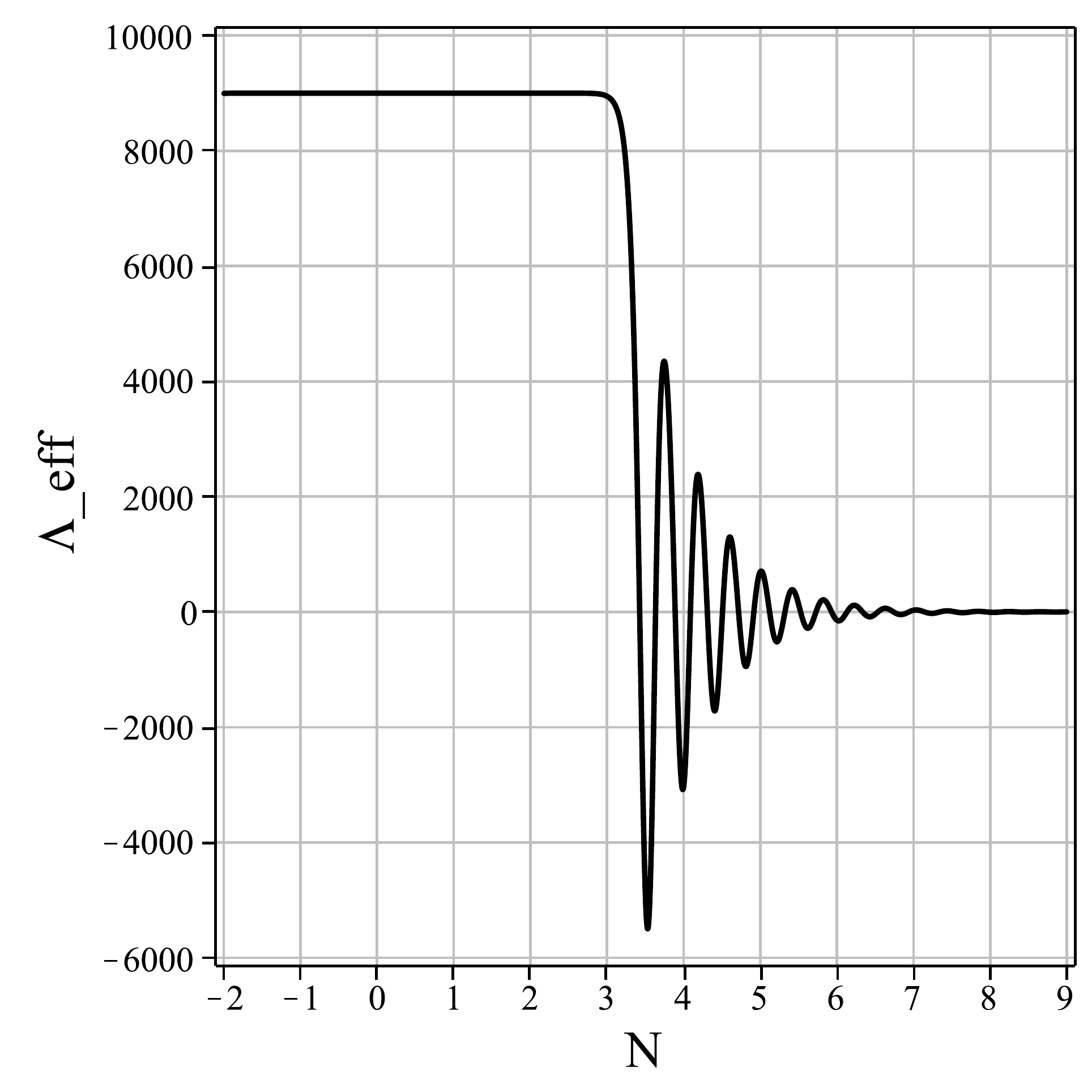} 
  \caption{Plot of $\phi(N)$ (left) and of $\Lambda_{\rm eff}$ (right) defined in eq.\ \eqref{Leff}. }
  \label{fig2}
  \end{figure}

\noindent In this section we solve numerically the equations \eqref{eomN} and we check that the analytical results found above are consistent. First, we confirm the stability character of the fixed points and then we show that there are trajectories that connect an inflationary Universe to a graceful exit with a reheating phase. We choose the values  $\lambda=1/10$ and $\xi=15$ and we consider for definitiveness only positive values of $z$ (i.e. of $\phi)$. In fig.\ (\ref{fig1}) (left) we plot the section $[H(N),\phi(N)]$ of the phase portrait of the full system of equations.  We choose the initial point of the trajectory at $N=0$ close to the unstable fixed point, by setting $x(0)=1$, $y(0)=z(0)=w(0)=10^{-8}$, and we let run the computation for 20 e-foldings. We see that the trajectory runs away from the initial point and spirals towards the stable fixed point  at $H(20)\simeq 0.7$, $\phi(20)\simeq 17$, consistently with the second of the relations \eqref{zeros}. In fig.\ (\ref{fig1}) (right) we plot the evolution of the Hubble parameter and we see a plateau followed by an oscillating phase. The same behaviour occurs for $\phi(N)$ as shown in fig.\ (\ref{fig2}) (left). In fig.\ (\ref{fig2}) (left) we plot instead the ``effective'' cosmological constant, defined as
\bea\label{Leff}
\Lambda_{\rm eff}={\alpha R^{2}\over 36}-{\lambda\phi^{4}\over 4}\,,
\eea   
and we verify that it vanishes as the Universe approaches the stable fixed point. Of course, the true evolution of the Hubble parameter after the oscillating phase is expected to be ruled by the matter fluid created via preheating, so these plots are no longer realistic after the first few oscillations of $H(N)$.


\section{Inflation}\label{sec4}


\noindent The inflationary phase of this model can be identified with the plateau clearly visible in the plot on the right in Fig.\ \ref{fig1}. We can give an approximate analytic description of this phase in the following way. From the first of eqs.\ \eqref{eomN} we can write the expression of the first Hubble flow parameter (expressed as a function of  $N$) that reads
\bea\label{eps1}
\epsilon_{1}\equiv -{H'\over H}={H^{2}\phi''+3H^{2}\phi'+\lambda \phi^{3}-4\xi\phi H^{2}\over H^{2}(\phi'-2\xi\phi)}\,.
\eea
Our goal is to constraint the parameters in such a way that inflation lasts a sufficiently long time. Conventionally, the end of inflation is marked by the time $N_{e}$ at which $\epsilon_{1}=1$ while, for $N<N_{e}$, $\epsilon_{1}<1$. The unstable fixed point is characterised by an arbitrary value of $H$ and vanishing $\phi$.  From eqs.\ \eqref{unsteq}, we know that, around the fixed point, we can approximate
\bea
H=H_{i}\,,\quad \phi\sim \phi_{i}\exp (N-N_{i})\,,
\eea
where  the subscript $i$ indicates a quantity  evaluated at the beginning of inflation. We neglect the exponentially decreasing part of $H(N)$ as it is irrelevant for the calculation. By replacing these expressions into eq.\ \eqref{eps1}, we find that $\epsilon_{1}=1$ when
\bea\label{deltan}
\Delta N\equiv N_{e}-N_{i}=\frac12\ln\left[(2\xi-3)H_{i}^{2}\over \lambda\phi_{i}^{2}\right]\,.
\eea
If we assume, as above, that $\xi\sim 1$ and that $\lambda \sim 10^{-8}$ we find that
\bea\label{efoldn}
{H_{i}\over \phi_{i}}\simeq \exp(\Delta N-9)\,,
\eea
which roughly fixes the necessary condition that guarantees $\Delta N=50 - 60$ e-folding of inflation.

These rough estimates are sufficient to show that there is an infinite number of inflationary trajectories springing from any point in the phase space close enough to the unstable fixed point in the Jordan frame. All these trajectories eventually end up at the stable fixed point where a mass scale emerges. Among these, only the ones satisfying the constraint \eqref{efoldn} are suitable to describe our Universe.


\section{Reheating}\label{sec5}


\noindent From the results in the previous section we learn that the model \eqref{fulllagra}  can describe an inflationary phase followed by a damped oscillations of the Hubble parameter and of the scalar field around a stable fixed point that breaks scale invariance and sets a mass scale. In this section we wish to study closely the oscillating phase and verify whether it can provide a reheating mechanism. This is necessary in order to connect the inflationary Universe to a radiation-dominated phase.

By using the explicit expression \eqref{oscil} for $H(N)$ close to the fixed point we  find that the scale factor averaged over several oscillation evolves as $a\sim t^{2/3}$. This indicates that the Universe expand as it was dominated by non-relativistic matter therefore we need some other mechanism to heat up the post-inflationary Universe into radiation-domination. We now explore some of the possibilities.

\subsection{Old reheating scenario}

\noindent The ``old reheating'' model is based on the assumption that the scalar field can decay into boson pairs $\chi$ (a pale simulacrum of the standard model fields) with a decay rate inversely proportional to the inflaton mass (for reviews on reheating see \cite{reviewreh}). The Lagrangian \eqref{fulllagra} must therefore  be augmented with new terms related to $\chi$ which, in the minimally coupled case, read
\bea\label{Ltot}
{\cal L}_{{\rm tot}}={\cal L}_{\rm inv}-g^{2}\phi^2\chi^{2}-\frac12(\partial \chi)^{2}-\frac12 m_{\chi}^{2}\chi^{2}\,,
\eea
where $g$ is the dimensionless coupling and $m_{\chi}$ is the mass of the decay product field $\chi$. Note that the scalar field $\phi$ is kept massless as a relic of the overall scale invariance of the initial action ${\cal L}_{\rm inv}$. However, nearby the stable fixed point, the scalar field oscillates around the equilibrium value $\phi_{0}$. Thus, we can expand the Lagrangian \eqref{Ltot} around $\phi_{0}$ upon the replacement $\phi\rightarrow \phi-\phi_{0}$. The relevant terms for the decay $\phi\rightarrow \chi+\chi$ are given by
\bea
{\cal L}_{{\rm reh}}\simeq -\frac12 m_{\phi}^{2}\phi^{2}+2g\phi_{0}\phi\chi^{2}+\ldots\,,
\eea
where $m_{\phi}$ is the effective mass for the inflaton that reads
\bea
m_{\phi}^{2}={3\lambda\phi_{0}^{2}\over 2}-{\xi \langle R\rangle\over 3}\simeq {\lambda\phi_{0}^{2}\over 2}\,.
\eea
where we replaced $R$ with its average value $\langle R\rangle =12H_{0}^{2}$ and we used the second of eqs.\ \eqref{zeros}.  The decay rate  can be estimated  by the quantity
\bea
\Gamma={g^{2}\phi_{0}^{2}\over 8\pi m_{\phi}}= \sqrt{2\over \lambda}{g^{2}\phi_{0}\over 8\pi}\,,
\eea
and, in the case when $m_{\chi}\ll m_{\phi}$ the decay $\phi\longrightarrow \chi+\chi$ is possible and transfers the energy stored in the field $\phi$ into the gas of relativistic particles $\chi$. The process stops at a time $H\sim \Gamma$ when the gas can finally thermalise at the reheating temperature
\bea
T_{\rm reh}\simeq \sqrt{\Gamma M_{p}}\simeq 0.3\times gM_{p} (\lambda \xi)^{-1/4}\,,
\eea
where we used the relation \eqref{mplanck}.

\subsection{Preheating}

\noindent A conspicuous particle production is possible via  parametric resonance of a scalar field $\chi$ coupled to $\phi$ in a way similar to the preheating scenario \cite{reheat}. Let us consider once more the Lagrangian \eqref{Ltot} and let us compute the Klein-Gordon equation associated to $\xi$, by neglecting for simplicity the mass term $m_{\chi}$.
In terms of Fourier modes, such equation can be written in the standard form
\bea\label{KGeq}
\ddot\chi_{k}+3H\dot\chi_{k}+\left({k^{2}\over a^{2}}+g^{2}\phi^{2}\right)\chi_{k}=0\,,
\eea
which describes an oscillator with the time dependent frequency
\bea
\omega_{k}=\left({k^{2}\over a^{2}}+g^{2}\phi^{2}\right)^{\frac12}\,.
\eea
In standard preheating, one defines the adiabaticity parameter
\bea
{\cal A}=\left| \dot\omega\over \omega^{2} \right|\,,
\eea
which characterise particle production. In general, whenever ${\cal A}\ll 1$ the production rate is negligible.  When this condition does not hold anymore, adiabaticity is broken and particle production can become effective. In the present case, we have
 \bea\label{adiab}
{\cal A}\simeq \left| {\phi' H\over g\phi^{2}} \right|\,.
\eea
where we have considered long-wavelength only, i.e. modes with with $k/(aH)\ll 2\pi$. We see immediately that the adiabaticity condition is broken when $\phi$ approaches zero. 

However, as we will shortly see, there is another particle production regime, which occurs when the oscillating function $H$ periodically vanishes around the fixed point. In this case, the adiabaticity condition, expressed as ${\cal A}\ll 1$, is not violated but  particle creation can still occur. This situation arises because there are more degrees of freedom as in usual preheating. If we look at the Einstein frame we basically have two scalar fields (one corresponding to the usual inflaton and the other associated to the quadratic term in $R$) that can amplify coupled light fields (for preheating in multified inflationary scenarios see \cite{kaiser, watanabe}).

Since we know explicitly the functions $H(N)$ and $\phi(N)$ around the stable fixed point (see eqs.\ \eqref{oscil}), we can solve equation \eqref{KGeq} almost exactly. We first define the new function
\bea
X_{k}=a^{3/2}\chi_{k}\,,
\eea
in terms of which we can express the comoving k-th bosonic occupation number 
\bea\label{nk}
n_{k}={\omega_{k}\over 2}\left( {|\dot X_{k}|^{2}\over \omega_{k}^{2}}+X_{k}^{2} \right)-\frac12\,,
\eea
which is obtained by inverting the usual formula for the harmonic oscillator (with $\hbar =1$)
\bea
E_{k}={\hbar \omega_{k}\over 2}(2n_{k}+1)=\frac12\left(|\dot X_{k}|^{2}+\omega_{k}^{2}|X_{k}|^{2}\right).
\eea
Then, eq.\ \eqref{KGeq} becomes, in terms of $N$-derivatives
\bea\label{Xeq}
X_{k}''+{H'X_{k}'\over H}+\left({k^{2}\over a^{2}H^{2}}+{g^{2}\phi^{2}\over H^{2}}\right)X_{k}=0\,.
\eea
Let us now consider eqs.\ \eqref{oscil}. We neglect the fast decaying part (i.e. we set $c_{2}=0$) and we consider $K\gg 1$. As a result we can approximate these functions as 
\bea\label{Hphi}
H(N)&=&H_{0}+T(N)\,e^{-\frac32 N}\,,\\\non
\phi(N)&=&\phi_{0}-\gamma\, T'(N)e^{-\frac32 N}\,,
\eea
where $H_{0}$ and $\phi_{0}$ are related by the second of eqs.\ \eqref{zeros} and where
\bea
T(N)=c_{3}\sin (KN)+c_{4}\cos (KN)\,,\quad \gamma={\xi\over 1+2\xi}\sqrt{\xi\over\lambda}\,.
\eea
If we further assume that, at least during the first oscillations, $H_{0}\ll T(N)$ and $\phi_{0}\ll\gamma T'(N)$, then eq.\ \eqref{Xeq} can be written as
\bea\label{eqz}
X_{k}''+\left({T'\over T}-\frac32\right)X_{k}'+\left(P^{2}+g^{2}\gamma^{2}{T'^{2}\over T^{2}}\right)X_{k}=0\,.
\eea
where we consider $P=k/(aH)\ll 1$ and  constant. This equation can be solved in two distinct physical regimes, both yielding particle production.

\subsection{$\phi$ - amplification}

\noindent We have seen that the condition \eqref{adiab} is violated when $\phi$ vanishes. This corresponds to $T'(N)\rightarrow 0$ while $T(N)$ is finite. In this regime, which occurs periodically,   eq.\ \eqref{eqz} simplifies into
\bea
X''_{k}-{3\over 2}X'_{k}+P^{2}X_{k}\simeq 0\,.
\eea
which can be solved with a linear combination of the modes
\bea
X_{k}\sim \exp \left(\frac34\pm\frac14\sqrt{9-16P^2}\right)\,,
\eea
as general solution. With these, it follows immediately that the comoving number mode evolves as
\bea
n_{k}\sim {T(N)\over P}\exp\left(\frac 32 N\right)\,,
\eea
every time $N$ is close to the critical value for which $T'(N)$ vanishes. Therefore, we recover the standard preheating picture of a periodic burst of particles every time the adiabaticity condition is violated. Note that this result is independent of the sign of $g^{2}$, so it can occur also for tachyonic couplings.

\subsection{$H$ - amplification}

\noindent In alternative to  standard preheating that occurs when $\phi(N)$ vanishes, there can be another particle production mechanism when $H(N)$ vanishes. In this regime in fact we can expand eq.\ \eqref{eqz} as
\bea
X''_{k}-{1\over N-N_{0}}X'_{k}+\left(P^{2}+{g^{2}\gamma^{2}\over (N-N_{0})^{2}}\right)X_{k}\simeq 0\,,
\eea
where $N_{0}=-{1\over K}\arctan (c_{4}/c_{3})$. This is a standard equation that can be solved in terms of Bessel's functions
\bea
X_{k}(z)=z\left[b_{1}J_{\nu}(Pz)+b_{2}Y_{\nu}(Pz)\right]\,,
\eea
where $z\equiv N-{N_{0}}$, $b_{1,2}$ are integration constants and $ \nu=\sqrt{1-g^{2}\gamma^{2}}$. In the limit of small $z$, the second term becomes dominant so \cite{grad}
\bea
X_{k}(z)\sim z^{1-\nu}\,,
\eea
which becomes very large for small $z$ when $1<\nu$, namely when $-g^{2}>0$. It is not hard to see that the same happens for the comoving particle number $n_{k}$, see eq.\ \eqref{nk}. This means that we can have particle production in the regime when $H(N)$ approaches zero while $\phi(N)$ is finite, provided the interaction term in ${\cal L}_{\chi}$ has $g^{2}<0$, as in the tachyonic preheating mechanism \cite{greene}.

In summary, according to the sign of $g^{2}$, we have two distinct particle amplification mechanisms. The first is the usual preheating model, which occurs when the oscillation of the field $\phi$ is no longer adiabatic, and it is valid for either signs of $g^{2}$. The second occurs when $g^{2}$ is negative as in tachyonic preheating, which implies ``negative'' effective square masses in the action but poses no stability problems if the potential is bounded from below.


\section{Analysis in the Einstein frame}\label{sec6}


\noindent We now come back to the inflationary phase of the model. As mentioned above, the analysis of this case is more transparent in the Einstein frame. The Lagrangian \eqref{fulllagra} can be written in the form
\bea\label{eqlagra}
{{\cal L}\over \sqrt{g}}=\chi R-{\alpha \varphi^{2}\over 36}-\frac12(\partial\phi)^{2}-{\lambda\over 4}\phi^{4}\,,
\eea
where we define the auxiliary variable \cite{wands}
\bea
\chi={\alpha\varphi\over 18}+{\xi \phi^{2}\over 6}\,.
\eea
The variation with respect to $\varphi$ vanishes only for $\varphi=R$ thus we can transfer the extra degree of freedom embedded in the $R^{2}$ term by formally introducing in \eqref{eqlagra} the field 
\bea
\varphi\equiv {18\chi\over \alpha}-{3\xi\phi^{2}\over \alpha}\,,
\eea
to obtain the equivalent Lagrangian
\bea\label{LagEq}
{{\cal L}\over \sqrt{g}}=\chi R-\frac{1}{2}(\partial\phi)^2-\frac{9}{\alpha}\left(\chi-\frac{\xi}{6}\phi^2\right)^2  = \chi R-\frac12(\partial\phi)^{2}-{\lambda\over 2}\phi^{4}+{3\lambda \chi\phi^{2}\over \xi}-{9\lambda \chi^{2}\over \xi^{2}}\,,
\eea
where we have used the relation \eqref{alpha}. This form describes two coupled scalar fields still in the Jordan frame. Note that $\chi$ has canonical mass-dimension two, just like the Brans-Dicke scalar, and that the action still enjoys the rigid Weyl symmetry scaling defined by $\chi'(x)=L^{-2}\chi(x)$ along with $\phi'(x)=L^{-1}\phi(x)$ and $g'_{\mu\nu}(x)=L^2g_{\mu\nu}(x)$. 

To study the dynamics in the Einstein frame we apply the conformal transformation $\tilde g_{\mu\nu}=\Omega^{2}g_{\mu\nu}$ where  $\Omega^{2}=2\chi/M^{2}$. As a word of caution, we note that the mass scale $M$ is arbitrary and it is inserted uniquely for dimensional consistency. In particular, we note that under the rigid Weyl symmetry scaling above, $\Omega$ is invariant if, and only if,  $M\to L^{-1}M$. As we will emphasise below,  $M$  has no relation with the breaking of scale invariance, which is still there, nor with any dynamically generated mass in the theory. 

 If we further define the field
\bea
\psi=\sqrt{6}M\ln \Omega\,,
\eea
the Einstein frame Lagrangian become
\bea
{\cal L}_{E}=\sqrt{\tilde g}\left[ {M^{2}\over 2}\tilde R-\frac12\tilde g^{\mu\nu}\partial_{\mu}\psi\partial_{\nu}\psi-\frac12\exp\left(-{\sqrt{2}\psi\over\sqrt{3}M}\right)\tilde g^{\mu\nu}\partial_{\mu}\phi\partial_{\nu}\phi-V(\phi,\psi) -{9\lambda M^{4}\over 4\xi^{2}}\right],
\eea
where $\tilde R$ is the Ricci tensor in the Einstein frame (see appendix D of \cite{wald}) and where
\bea
V(\phi,\psi)={\lambda\phi^{4}\over 2}\exp\left(-{2\sqrt{2}\psi\over\sqrt{3}M}\right)-{3\lambda M^{2}\phi^{2}\over 2\xi}\exp\left(-{\sqrt{2}\psi\over\sqrt{3}M}\right)\,.
\eea
The Lagrangian is not yet canonical because of the factor that multiplies the kinetic term of $\phi$. Eventually, this term can be written in canonical form provided one introduces the new scalar field defined by the differential relation
\bea
d\tilde\phi = \Omega^{-1}d\phi\,,
\eea
but, for our purposes, this is not necessary.

To study the equations of motion, we find convenient to define the following quantities
\bea
f(t)&=&M\exp\left(-{\sqrt{2}\psi(t)\over 2\sqrt{3}M}\right)\,,\\\non
\Lambda&=&{9\lambda M^{2}\over 4\xi^{2}}\,,\\\non
V&=&f^{2}\phi^{2}(-q_{1}+q_{2}f^{2}\phi^{2})\,,\quad q_{1}={3\lambda\over 2\xi}\,,\quad q_{2}={\lambda\over 2M^{4}}\,.
\eea
The Lagrangian finally reads
\bea\label{Elagra}
{\cal L}_{E}=\sqrt{g}\left[ {M^{2}\over 2}(R-2\Lambda)-{3M^{2}\over f^{2}}(\partial  f)^{2}-{f^{2}\over 2M^{2}}(\partial \phi)^{2}-V  \right]\,,
\eea
where we dropped the tilde to simplify notation.
The Friedmann equations, with a flat Robertson-Walker metric are
\bea
H^{2}&=&{\dot f^{2}\over f^{2}}+{f^{2}\dot \phi^{2}\over 6M^{4}}+{\Lambda\over 3}+{V\over 3M^{2}}\,,\non\\
\dot H&=&-{3\dot f^{2}\over f^{2}}-{f^{2}\dot\phi^{2}\over 2M^{4}}\,,
\eea
while the Klein-Gordon equations for the two scalar fields are
\bea
&&\ddot \phi+3H\dot \phi +{2\dot \phi \dot f\over f}+{M^{2}\over f^{2}}{\partial V\over \partial \phi}=0\,,\non\\
&&\ddot f+3H\dot f -{\dot f^{2}\over f}-{f^{3}\dot \phi^{2}\over 6M^{4}}+{f^{2}\over 6M^{2}}{\partial V\over \partial f}=0\,.
\eea
We stress once again that scale invariance is still present in the equations above, despite the appearance of a the mass scale $M$. In fact, the equations of motion are invariant under the (actively interpreted) scale transformations 
\bea
\bar H(t)=\ell H(\ell t)\,,\quad \bar \phi(t)=\ell \phi(\ell t)\,,\quad \bar f(t)=\ell f(\ell t)\,,\quad \bar M= \ell M\,,
\eea
consistently with the transformations \eqref{scinv2}. The last transformation may seems a trivial change of units, but actually it is not since $M$ is the scale of a dynamical field, which is not neutral under dilatations. To be more precise, since dilatations (both in classical theories as well as in their quantum version) map the square of the four momentum $P^2$ into $\ell^{-2}P^2$, scale invariance requires that all masses either vanish or form a continuous spectrum, but says nothing about parameters like $M$ that are introduced only for dimensional purposes and are not part of the mass spectrum (after all, the scalar field itself is a mass from the dimensional analysis point of view, although its scale is arbitrary anyway).  

As it will be apparent below, all the observables of interest here (such as number of e-folds and spectral indices) are independent of $M$, which reflects the fact that $M$ is a so-called redundant parameter, as we explicitly show in the Appendix. For these reasons, in the following we keep considering $M$ as an arbitrary mass scale.

The analysis of the dynamical system associated to the equations above, analogous to the one made in the Jordan frame, confirms that there are two fixed points. As before, one is unstable and  located at
\bea
H_{\rm unst}={\sqrt{3 \lambda} M\over 2\xi}\,,\quad \phi_{\rm unst}=0\,,\quad f_{\rm unst}={\rm arbitrary}\,,
\eea
for arbitrary $f_{\rm unst}$. The instability is confirmed by solving the linearised system around this point, which yields $\psi\sim t$, $\phi\sim \pm \sqrt{3\lambda/\xi}\, M\,t$, revealing that both fields have growing modes.

The second fixed point is stable and located in 
\bea
H_{\rm stab}={\sqrt{3\lambda}M\over 2\sqrt{2}\xi}\,,\quad \phi_{\rm stab}=\phi_{0}={\rm arbitrary}\,,\quad f_{\rm stab}={\sqrt{3}M^{2}\over \sqrt{2\xi}\phi_{0}}={M\over \sqrt{2}}\,,
\eea
where, for the last identity, we used the relation \eqref{mplanck} and the fact that the $\phi$-coordinate of the unstable fixed point is unchanged upon the conformal transformation. The general solution of the linearised system contains oscillating, decreasing and growing modes but, since this fixed point must also have $\dot f=\dot\phi=0$, the latter are excluded from the spectrum. Therefore, the physical solution is a combination of decaying (and oscillating) modes only.
Note that, as in the Jordan frame,  the potential  and the cosmological constant term cancel each other  at the stable fixed point. Note also that the ratio
\bea\label{ratio}
{H_{\rm unst}\over H_{\rm stab}}=\sqrt{2}\,,
\eea
is independent of the parameters. 

An observation is in order here. The relation between the Hubble parameter in the Jordan frame ($H_{J}$) and the one in the Einstein frame ($H_{E}$) is given by $H_{J}=\Omega H_{E}$ \cite{chiba}. One can easily check that this formula holds at the stable fixed point, where $\Omega_{\rm stab}=\sqrt{2}$. For the unstable one it holds as well provided one recognises that the conformal factor at the unstable point depends on $f_{\rm unst}$, which is arbitrary. This is why $H_{\rm unst}^{J}$ is arbitrary while $H_{\rm unst}^{E}$ is not.

After these general considerations, let us focus on the inflationary solutions. However, it is important to clarify first that $H_{\rm unst}$ \emph{does not} characterise the inflationary value of the Hubble parameter in the Einstein frame. In fact, the inflationary trajectories are uniquely determined by the condition $|\phi|=|\dot \phi|\ll 1$, which is the same as in the Jordan frame (we stress once again that the scalar field is the same in the two frames). Therefore, eq.\ \eqref{ratio} is not the ratio between the current Hubble parameter and the inflationary one. With this in mind, we consider $\phi$ very small and constant, so we can neglect the term $(\partial\phi)^{2}$, and write the Lagrangian \eqref{Elagra} as 
\bea\label{leff}
{\cal L}=\sqrt{g}\left[{M^{2}\over 2}R  -\frac12 (\partial\psi)^{2}-W \right]\,,
\eea
where 
\bea
W={\lambda\phi^{4}\over 2}\exp\left( -{2\sqrt{2}\psi\over \sqrt{3}M} \right)-{3\lambda M^{2}\phi^{2}\over 2\xi}\exp\left(- {\sqrt{2}\psi\over \sqrt{3}M} \right)+{9\lambda M^{4}\over 4\xi^{2}}\,.
\eea
We see that the effective action is quite similar to the Starobinsky model written in the Einstein frame. In fact, we now show that the model \eqref{leff} predicts the same spectral indices, at least to the first order in the slow-roll parameters. To see this, we compute 
\bea\label{slowroll}
\epsilon={M^{2}\over 2}\left({1\over W}{\partial W\over \partial \psi}\right)^{2}\,,\quad \eta={M^{2}\over W}{ \partial^{2}W\over \partial \psi^{2} }\,,
\eea
and the scalar spectral index and the tensor-to-scalar ratio, given respectively by
\bea\label{indices}
n_{s}=1+2\eta-6\epsilon\,,\quad r=16\epsilon\,.
\eea
Since  $\phi$ is constant, we can eliminate it by combining $n_{s}$ and $r$. The resulting expression is quite involved but  can  be expanded for $n_{s}\rightarrow 1$ giving
\bea
r\simeq 3(n_{s}-1)^{2}\,,
\eea
which is the same expression found for the Starobinsky model. Let us now define the function
\bea\label{efold}
N={1\over M^{2}}\int d\psi\Bigg|W\left(\partial W\over \partial \psi\right)^{-1}\Bigg| \,.
\eea
The number of e-foldings between some initial value of $\psi_{i}$ and the end of inflation at $\psi_{f}$ is defined by
\bea
N^{\star}=N(\psi_{i})-N(\psi_{f})\,.
\eea
The value $\psi_{f}$ is conventionally set at the earliest time at which $\epsilon=1$ or $|\eta|=1$. In our case we find that $\epsilon=1$ occurs before $|\eta|=1$ and that $N(\psi_{f})\simeq 0.78$. Since inflation must last at least $N^{\star}=50$ e-foldings we can safely neglect $N(\psi_{f})$. At last, since during inflation (i.e. near the unstable fixed point), we also have  $\phi/M\ll 1$, we expand the expression for $\epsilon$ around $\phi=0$, we integrate eq.\ \eqref{efold}, we identify $N^{\star}=N(\psi_{i})$  and we find
\bea
N^{\star}\simeq {9 M^{2}\over 4\xi\phi^{2}}\exp\left(\sqrt{2}\psi_{i}\over \sqrt{3}M \right).
\eea
If we insert this result in the expressions \eqref{slowroll} and \eqref{indices} to eliminate again $\phi$, we find
\bea
\epsilon\simeq {3\over 4(N^{\star })^{2}}\,,\quad \eta\simeq -{1\over N^{\star}}\,, \quad \rightarrow\quad n_{s}\simeq 1-{2\over N^{\star}}-{3\over 2(N^{\star})^{2}}\,.
\eea
These results coincide with the prediction of the Starobinsky model, independently of the the value $\psi_{i}$ and with the only condition  $\phi/M\ll 1$, i.e that inflation begins close enough the unstable fixed point of the system.

As anticipated above, these results are independent of the choice of $M$, which is arbitrary in light of the intrinsic scale invariance of the model. We believe that a more sophisticated analysis of the perturbations, which goes beyond the scope of this paper, can determine to what measure this model differs from the original Starobinsky model, see e.g. \cite{Qiu:2014apa} for  suitable techniques.


\section{Conclusion}\label{sec7}


\noindent In this paper we have examined a simple theory based on a scale-invariant Lagrangian with quadratic gravity and a non-minimally coupled scalar field. Scale invariance spontaneously breaks when the field configuration approaches a stable fixed point, where a mass scale emerges. The latter can be identified with the Planck mass but the inflationary observables are independent of such a choice. In fact,  as there are infinitely many minima, the precise choice of the value of $M$ has to be made by hand or possibly via anthropic arguments. The model as it is cannot predict the value of the effective Planck mass. If no other fields are introduced into the model, the perturbations around the fixed point are still governed by scale invariant field equations, with the symmetry realized non linearly (the symmetry is broken in the vacuum, not in the Lagrangian). However this is unrealistic: according to standard lore, during reheating the standard model fields get excited and even perturbations breaks the original scale invariance, which is lost forever.

The global evolution of the system brings the Universe from an inflationary phase to a graceful exit, when the Universe reheats through various mechanisms that have been reviewed. From the phenomenological point of view, the inflationary predictions of this model are the same of Starobinsky's inflation but, in contrast to the latter, there is no need to introduce a second mass scale.

An interesting feature is that, in principle, the model depends only on two parameters, namely the strength of the non-minimal coupling of the scalar field to gravity and  the strength of the scalar quartic self-interaction. However, the inflationary predictions are in fact independent of these parameters, at least at the leading term, and this reflects again the underlying scaling symmetry. 

The dynamical evolution of this system towards the stable fixed point at a  non-zero value of the Hubble parameter can be interpreted as the existence of a relic cosmological constant, which may be compatible with current observations at the price of an extreme fine tuning of the scalar self-coupling. In this case one should remember that the effective value of the cosmological constant can be affected by several other kinds of contributions, including the vacuum energy of quantum fields or the classical dynamics of Yang-Mills fields \cite{me}. Alternatively, it can be considered as the initial value of the following radiation era, in which case it  will start to decrease in the usual way from a value around $10^{14}\,\rm{Gev}$ to the present, but the late time acceleration has to be obtained by other means.

There are several aspects that are left for future work. First of all, the weight of quantum corrections has not been computed here. In fact, it may be possible that these alter the stability configuration of the system or the inflationary predictions. Another aspect to be studied is preheating in the Einstein frame, where there are two scalar fields interacting. At the classical level, these fields undergo damped oscillation but, at the quantum level, they might interact and produce other particles through resonant amplifications. In addition, the inflationary dynamics is essentially determined by both scalar fields, although the fixed point analysis justifies the choice of a very small and constant $\phi$ made in the previous section. In general, it would be interesting to study the model also when these conditions on $\phi$ are relaxed.  Finally, the theory presented here is minimal, in the sense that there are several other terms that are scale invariant and that could be added to the Lagrangian \eqref{fulllagra}. We believe that our model is sufficient to capture the main characteristics but it would certainly interesting to consider more general setups.

\appendix*
\section{Redundant parameter $M$}

\noindent After the metric rescaling $\tilde g_{\mu\nu}=\Omega^{2}g_{\mu\nu}$, where $\Omega^{2}=2\chi/M^{2}$ and $M$ is an arbitrary mass parameter, the action associated to the Lagrangian \eqref{LagEq} reads
\bea\label{acM}
S=\int d^{4}x\sqrt{g}\left[ {M^{2}R\over 2}-{3M^{2}\over \Omega^{2}}(\partial \Omega)^{2}-{\Omega^{2}\over 2} (\partial \phi)^{2} -{\lambda \phi^{4}\Omega^{4}\over 2}-{9\lambda M^{4}\over 4\xi^{2}}+{3\lambda\phi^{2}M^{2}\Omega^{2}\over 2\xi }\right]\,.
\eea
We now prove that $M$ is a redundant parameter, following the definition given in \cite{weinb}. The trace of the Einstein equations obtained from \eqref{acM} is
\bea\label{trace}
R-{6\over \Omega^{2}}(\partial \Omega)^{2} -{9\lambda M^{2}\over \xi^{2}}= {\Omega^{2}\over M^{2}}(\partial \phi)^{2}+{2\lambda \phi^{4} \Omega^{4}\over M^{2}}-{6\lambda \phi^{2} \Omega^{2}\over \xi}\,.
\eea
The Klein-Gordon equation for $\phi$ is
\bea\label{KGM}
{1\over \sqrt{g}}\partial_{\mu}\left(\sqrt{g}\Omega^{2}\partial^{\mu}\phi\right)=2\lambda\phi^{3}\Omega^{4}-{3\lambda \phi M^{2} \Omega^{2}\over \xi}\,.
\eea
By differentiating with respect to $M$ we find
\bea
{1\over M}{\partial S\over  \partial M} =\int d^{4}x\sqrt{g}\left[ R-{6\over \Omega^{2}}(\partial \Omega)^{2}  -{9\lambda M^{2}\over \xi^{2}}+{3\lambda \phi^{2}\Omega^{2}\over \xi} \right]\,.
\eea
By substituting eqs.\ \eqref{trace} and \eqref{KGM} into the above equation gives
\bea
M{\partial S\over  \partial M} = \int d^{4}x\sqrt{g}\,\Omega^{2}(\partial\phi)^{2}+\int d^{4}x\,\phi\,\partial_{\mu}\left(\sqrt{g}\,\Omega^{2}\partial^{\mu}\phi\right)\,,
\eea
which vanishes upon integration by parts of the second term. This shows that $M$ is a redundant parameter since the variation of the action with respect to $M$ vanishes by using the field equations.

\begin{acknowledgments}
The authors would like to thank S.\ Zerbini, G.\ Cognola, and G.\ Tambalo for fruitful discussions.
\end{acknowledgments}


\end{document}